\documentclass[aps,prb,twocolumn,preprintnumbers,amsmath,amssymb]{revtex4}
\usepackage{graphicx}
\usepackage{xcolor}
\begin{document}
\title{Magnetoresistance and Kohler rule in the topological chiral semimetals CoSi.}

\author{A.~E.~Petrova}
\affiliation{P.~N.~Lebedev Physical Institute, Leninsky pr., 53, 119991 Moscow, Russia}
\author{O.~A.~Sobolevskii}
\affiliation{P.~N.~Lebedev Physical Institute, Leninsky pr., 53, 119991 Moscow, Russia}
\author{S.~M.~Stishov}
\email{stishovsm@lebedev.ru}
\affiliation{P. N. Lebedev Physical Institute, Leninsky pr., 53, 119991 Moscow, Russia}

\begin{abstract}
The transverse and longitudinal magnetoresistance (MR) of two samples of the topological chiral semimetal CoSi with different RRR was studied. It is shown that the Kohler rule  works for the transverse MR. The Kohler rule is also fulfilled in the case of longitudinal MR at a low reduced magnetic field.  A sharp deviation of longitudinal MR curve for sample with low RRR from the Kohler prediction at high fields reveals its tendency to a sign change at higher magnetic fields. The Shubnikov – de Haas quantum oscillations were observed and analyzed in both perpendicular and parallel configurations of the current and magnetic field in sample CoSi-1 with RRR=9.33 at low temperatures. 
\end{abstract}
\maketitle

\section{Introduction}

It is well known that for the majority of normal metals the transverse magnetoresistance (TMR) is proportional to $H^2$ in small fields, whereas the longitudinal magnetoresistance (LMR) is zero in the case of spherical Fermi surface. But in a general case, the LMR can be also significant and positive (see Ref.~\cite{Bla,Pip}). However, it has been recently proposed that in nonmagnetic Weyl semimetals the chiral anomaly leads to a novel kind of low field LMR: negative and quadratic in the magnetic field~\cite{Son,Bur}. Then the negative LMR phenomena were discovered in some topological chiral materials with the Weyl points in electron spectra Ref.~\cite{Xia,Chen,Qia}. So, the negative LMR may serve as a signature of the so-called chiral anomaly. The violation of the Kohler rule, which is a scaling law ($MR=f[H/\rho_0]$)~\cite{Bla,Pip} can also be one of the features of topological chiral materials~\cite{Jin}.

\begin{figure}[htb]
\includegraphics[width=80mm]{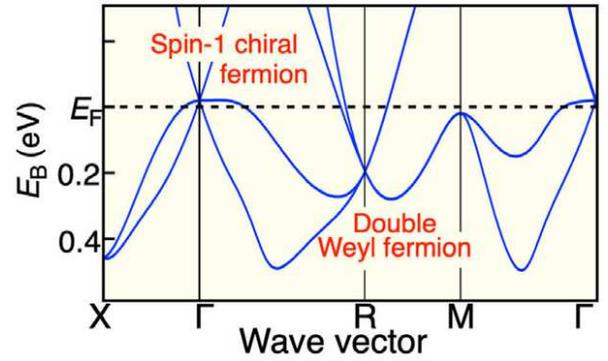}
\caption{\label{fig1} (Color online) Band structure of CoSi, redrawn from~\cite{Tak}. Spin-orbit splitting is not shown.}
\end{figure}

In this connection, the chiral semimetal CoSi is a subject of interest. CoSi belongs to the class of transition metal silicides and germanides with a non-centrosymmetric B20-type crystal structure. Interest in compounds of this kind has increased significantly due to their topological properties, which are determined by the specific symmetry of their crystal lattice. The violation of spatial symmetry in the B20 structure type causes the existence of special points, including Weyl points, in electronic and phonon spectra, characterized by topological Chern numbers (see Ref.~\cite{Arm,Van}). At the same time, semimetals such as CoSi and RhSi, and, possibly, CoGe and RhGe occupy a special position, revealing nontrivial fermions of various kinds ~\cite{Tan,Cha}.
CoSi is a diamagnetic semimetal with a very high residual resistance, which suggests the presence of a fairly large number of defects~\cite{Sti}. Currently, CoSi is being actively studied as a kind of model material demonstrating new types of Fermi quasiparticles (see Fig.~\ref{fig1}).
Note that the electronic properties of CoSi generally agree well with the calculations of the band structure and density of states, which are supported by the ARPES (see, for example, Ref.~\cite{Tak}).

Earlier the MR in CoSi was studied in Ref.~\cite{Sha,Xit,Wu}. As was found in Ref.~\cite{Sha,Xit,Wu}, a dependence of the MR on the magnetic field is described by the expression $MR\sim H^n$, where n varies from 1.6 to 1.8, which is different from the canonic n=2 and probably reveals a trend to saturation. The Kohler rule is shown to work in the case of polycrystalline sample of CoSi when the zero resistivity is varied by temperature. No samples of CoSi with different zero resistivity were studied~\cite{Sha}.
It seems that no negative resistivity features in bulk CoSi were found so far, though claims on negative resistivity in single-crystalline CoSi nanowires ~\cite{Seo} and CoSi thin films on silicon~\cite{Chi} should be mentioned. 
With all that in mind, we decided to study the MR of the chiral semimetal CoSi.

\section{Experimental} 
Two samples of CoSi were selected for the current study. Sample CoSi-1, grown by the Czochralski method, with resistivity $\rho=14.9\ \mu\Omega\ cm$ at 2~K has the size $\sim$0.4x1.1x9.5 mm$^3$. Sample CoSi-2, grown by the Bridgman technique, with resistivity $\rho=65.8 \ \mu\Omega\ cm$ at 2~K has the size $\sim$0.7x1.3x7.5 mm$^3$. Lattice parameters of samples determined by powder x-ray diffraction are 4.444(1)~\AA\ (CoSi-1) and 4.443(1)~\AA\ (CoSi-2). Chemical analysis performed with an electron probe x-ray microanalyzer showed some deviation from the stoichiometric chemical composition (silicon deficiency 1\%). Of the impurities have been determined by atomic emission and mass spectroscopy a noticeable amount of Ni ($\sim$0.01\%) was detected. 
\begin{figure}[htb]
\includegraphics[width=80mm]{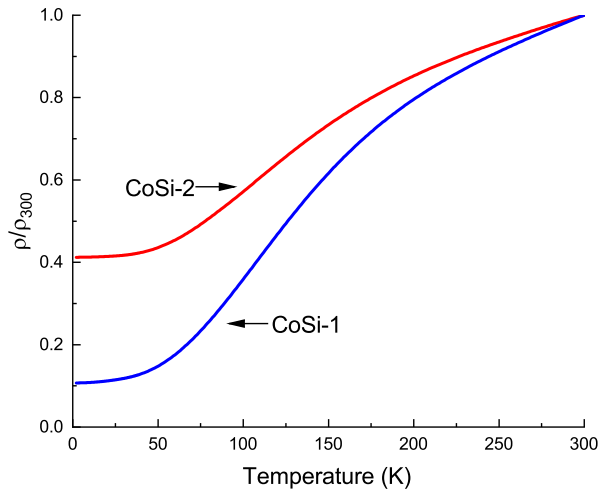}
\caption{\label{fig2} (Color online) Temperature dependence of resistivity of samples CoSi.}
\end{figure}
Electrical transport measurements in magnetic fields up to 16 T were conducted using Cryogen Free Measurement System (CFMS-16) from Cryogenic Ltd. Magnetoresistance was measured with a four-probe AC method. The magnetic field was applied in two different orientations: parallel and perpendicular to electrical current. Measurements were conducted in the following conditions: electrical current 500~$\mu$A, measurement frequency 175.38 and 313.3 Hz for each sample, peak-to-peak noise less than 40 nV, the current and voltage wires were twisted in pairs reducing noises in high magnetic fields. We use Standard Research Current Source 580 and Lock-In SR~830. Temperature was measured with the Cernox temperature sensor. Temperature stability was about 10~mK.

The resistivities of samples are shown in Fig.~\ref{fig2}. The high residual resistivity along with the low residual resistivity ratio (RRR) and tendency to saturation at high temperatures place these samples in the category of so-called strongly disordered metals, whose resistivity can be described by a parallel resistor model~\cite{Wie}.
\begin{figure}[htb]
\includegraphics[width=80mm]{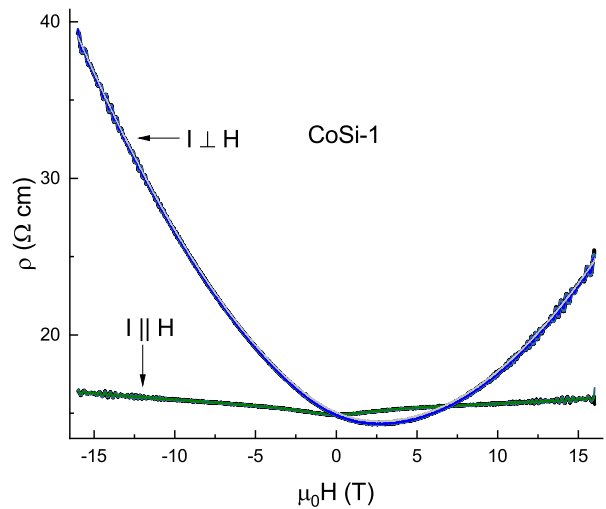}
\caption{\label{fig3} (Color online) Resistance of CoSi-1  as a function of the magnetic field.}
\end{figure}

Results of magnetoresistance measurements are illustrated in Fig.~\ref{fig3}.
As is seen in Fig.~\ref{fig3} the "transverse" data, which should be quadratic in the magnetic field look highly asymmetric as a difference to the longitudinal ones.

\section{Results and discussion}
\begin{figure}[htb]
\includegraphics[width=80mm]{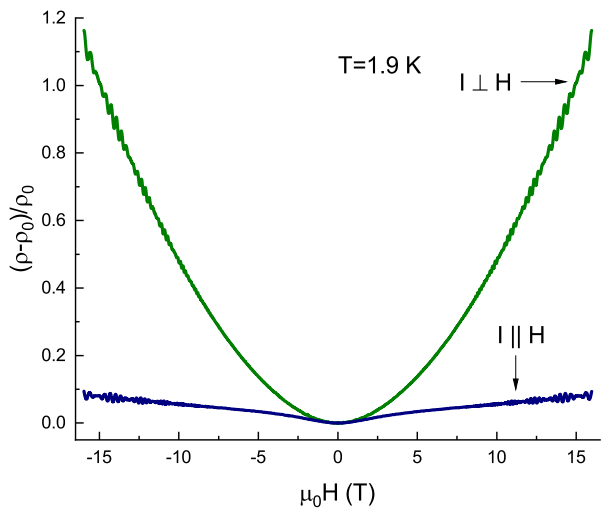}
\caption{\label{fig4} (Color online) Relative change of resistance of CoSi-1 in magnetic fields, corrected for the Hall contribution.}
\end{figure}

\begin{figure}[htb]
\includegraphics[width=80mm]{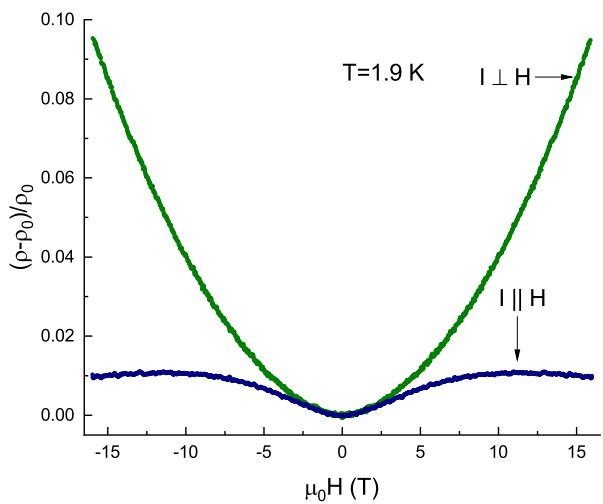}
\caption{\label{fig5} (Color online) Relative change of resistance of CoSi-2 in magnetic fields, corrected for the Hall contribution.}
\end{figure}

Naturally,  it occurs to us that the Hall component arising due to the sample misalignments in the magnetic field and not quite correct positions of the electrical contacts may be responsible for that. Taking into account that the MR is an even function of the magnetic field a simple data manipulation permits a subtraction the odd Hall contributions from  the raw experimental data, which results in  Figs.~\ref{fig4} and ~\ref{fig5}. These figures clearly demonstrate a lack of negative magnetoresistance in the LMR curves at least up to 15 T. However, as is seen in Figs.~\ref{fig4} and ~\ref{fig5}, a behavior of the LMR curves in magnetic fields is strongly different from the TMR ones. Moreover, the LMR for the high resistivity sample CoSi-2, (Figs.~\ref{fig5} and ~\ref{fig7}) supposedly may indicate a tendency to negative values of MR in much high magnetic fields. Note that the mentioned tendency is revealed in sample CoSi-2, which does not support the quantum oscillations. 
 
\begin{figure}[htb]
\includegraphics[width=80mm]{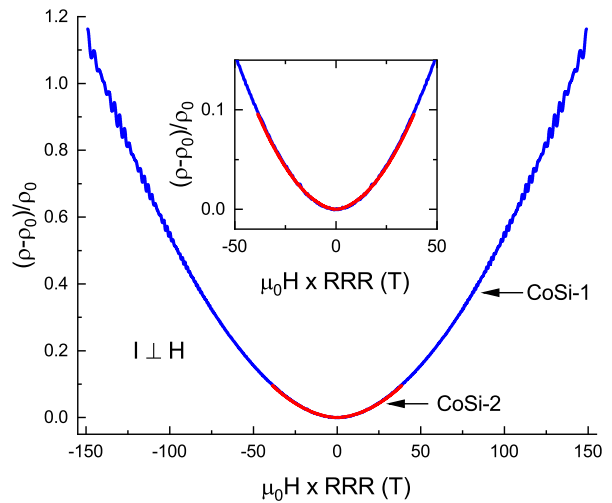}
\caption{\label{fig6} (Color online) Kohler plot for TMR of samples CoSi-1 and CoSi-2.}
\end{figure}

\begin{figure}[htb]
\includegraphics[width=80mm]{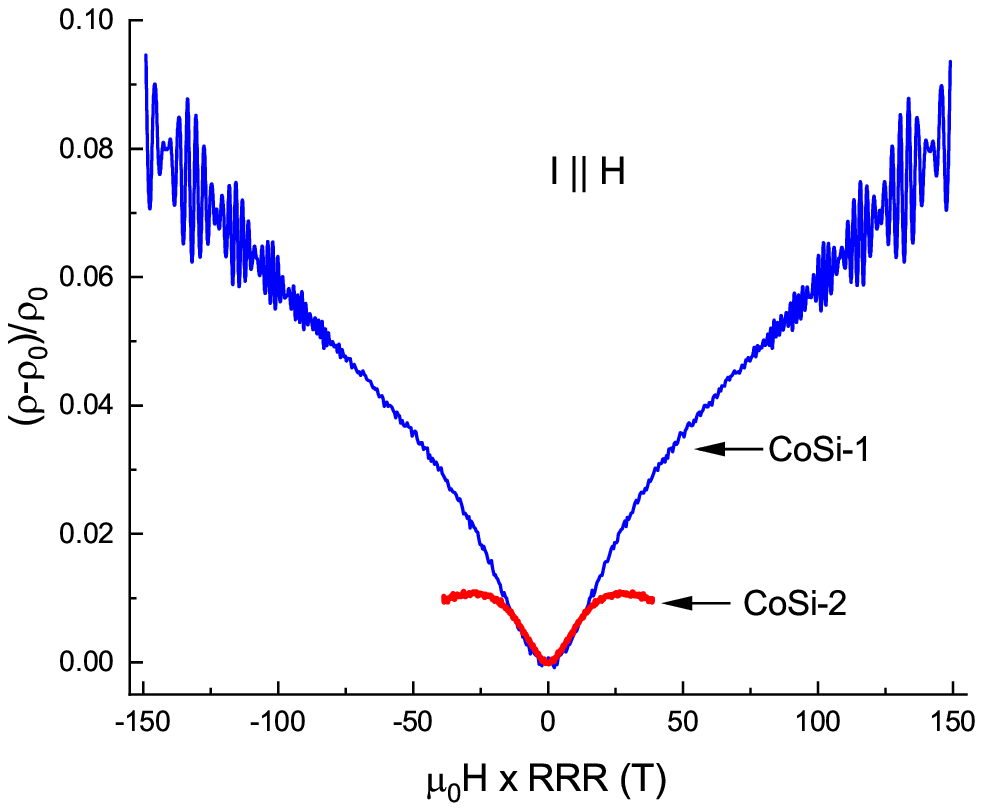}
\caption{\label{fig7} (Color online) Kohler plot for LMR of samples CoSi-1 and CoSi-2.}
\end{figure}

It should be pointed out that approximations of the  LMR data (Figs.~\ref{fig5},~\ref{fig6}) by the expression $MR\sim H^n$ give an overall value $n\cong1.8$ in some agreement with data [14-16]. Changes $n$ from the theoretical value 2 to 1.8  may be a result of upcoming saturation.

Now we turn to an analysis of the Kohler rule, which is written in a standard form as $\frac{\rho-\rho_0}{\rho_0}=F(\frac{H}{\rho_0})$ where $\rho_0$-resistivity at H=0, \textit{F}-function, which does not depend on temperature, but changed for different metals. A violation of the Kohler rule was observed for some topological materials in Ref.~\cite{Jin}. Authors of Ref.~\cite{Jin} suggested a modification of the Kohler rule in the form  $\frac{\rho-\rho_0}{\rho_0}=F(\frac{H}{n_T\rho_0})$, where coefficient $n_T$ supposedly accounts for a change of carrier density. 

 \begin{figure}[htb]
\includegraphics[width=80mm]{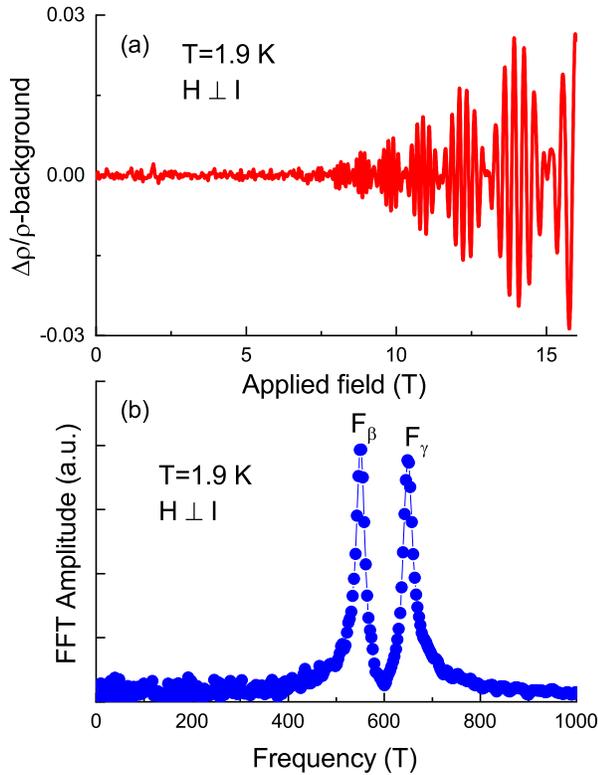}
\caption{\label{fig8} (Color online) Oscillating part of TMR and corresponding FFT spectra for $H\perp I$.  We borrowed the peak designations from Ref.~\cite{Wan}, where also
 the low frequency peak $F\alpha$, corresponding to $\Gamma$-point, was discovered.}
\end{figure}

The Kohler plots for two samples of CoSi with quite different resistivities (see Fig.~\ref{fig2}) are shown in Figs.~\ref{fig6} and \ref{fig7}. The Pippard approach was applied when RRR is used as a scaling coefficient~\cite{Pip}. As is evident, the Kohler rule nicely works for the transverse MR (Fig.~\ref{fig6}). Surprisingly, the Kohler rule is also fulfilled in the case of longitudinal MR at a low reduced magnetic field (Fig.~\ref{fig7}). At the same time, a sharp deviation of longitudinal MR curve for sample CoSi-2 from the Kohler prediction at high fields supports our conclusion about its tendency to a sign change at higher magnetic fields (see Figs.~\ref{fig5} and~\ref{fig7}). However, whether this feature is stipulated by the chiral anomaly remains to be seen in future studies.

	Finally, we cannot leave without attention the resistance oscillations in magnetic fields clearly seen in the Figs.~\ref{fig4},\ref{fig6},\ref{fig7}. These oscillations are a revelation of the Shubnikov–de Haas quantum effect and can be observed in the both perpendicular and parallel configurations of the current and magnetic field of sample CoSi-1 with RRR=9.33 at low temperatures. Sample CoSi-2 with RRR=2.42 does not support the quantum oscillations because the electron mean free path is too short. Fig.~\ref{fig8}a demonstrates the oscillating part of the TMR resulting from subtracting the background from the MR curve (Fig.~\ref{fig4}). The background was approximated by the "locally weighted polynomial regression" approach~\cite{Jek}. Two fundamental frequency peaks in the spectra arose after the Fast Fourier transform (Fig.~\ref{fig8}b). The peak positions were determined from the Lorentzian fits. Similar procedure was performed for the LMR (Fig.~\ref{fig4}) and two slightly different frequencies were obtained (see Table~\ref{table1}).

\begin{figure}[htb]
\includegraphics[width=80mm]{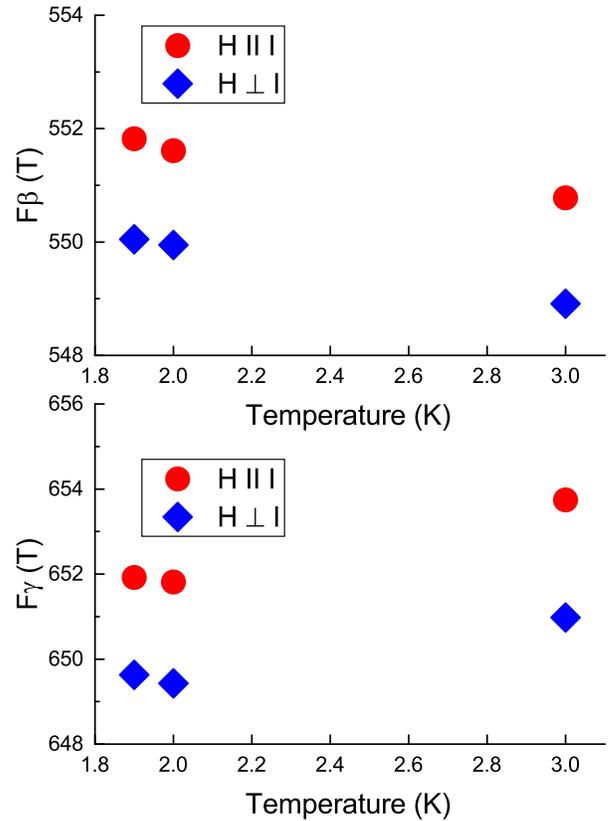}
\caption{\label{fig9} (Color online) Variations of the FTT peak frequencies with temperature. Errors of the peak positions are about 1-3T }
\end{figure}

The found frequencies correspond to the extremal cross section area of almost spherical electron pockets of the Fermi surface situated at $R$ point in the $BZ$ corner. These “spheres” are split by the spin-orbit coupling, which results in the observed two close frequencies Ref.~\cite{Xit,Wan}. The obtained data are shown in Table~\ref{table1}, where also placed some earlier results (Ref.~\cite{Xit,Wu,Wan}) for comparison purpose.
Variations of the FTT peak frequencies with the experimental setup and temperature are displayed in Fig.~\ref{fig9}. In connection with the calculational procedure, included some approximation, we cannot claim absolute reliability of our data in respect of the effects arising at change of the magnetic field direction and temperature. On the other hand, as was shown in the de Haas-van Alphen study the electron pockets in R point are not completely spherical~\cite{Wan}. Hence, the orientational dependence of the oscillating frequencies may be real. The temperature dependence of the oscillating frequencies in Fig.~\ref{fig9} implies that the spin-orbit splitting probably increases with temperature.

\begin{table}
\caption{\label{table1} The frequencies.}

\begin{tabular}{|c|c|c|c|c|}
\hline 
$F\beta\gamma$ & $1F\beta \perp$ & $1F\gamma \perp$ & $2F\beta \parallel$ & $2F\gamma \parallel$ \\ 
\hline 
Present data & 550.05 & 649.63 & 551.82 & 651.92 \\ 
\hline 
Ref.~\cite{Sha} & 671 & 671 & - & - \\ 
\hline 
Ref.~\cite{Xit} & 563.9 & 664.6 & 564.5 & 664.1\\ 
\hline 
Ref.~\cite{Wie} & 557.6 & 656.7 & - & - \\ 
\hline 
\end{tabular} 

\end{table}

\section{Conclusions}
The transverse and longitudinal magnetoresistance of two samples of the topological chiral semimetal CoSi with different RRR was studied. It is shown that the Kohler rule  works for the transverse MR. The Kohler rule is also fulfilled in the case of longitudinal MR at a low reduced magnetic field.  A sharp deviation of longitudinal MR curve for sample with low RRR from the Kohler prediction at high fields reveals its tendency to a sign change at higher magnetic fields. The Shubnikov – de Haas quantum oscillations were observed in both perpendicular and parallel configurations of the current and magnetic field in sample CoSi-1 with RRR=9.33 at low temperatures. Two fundamental frequency peaks in the spectra arose after the Fast Fourier transform applied to the oscillation parts of the curves in Fig.~\ref{fig4}, which is illustrated by Fig.~\ref{fig8}. The same procedure was performed for the LMR (Fig.~\ref{fig4}) and two slightly different frequencies were obtained (see Table~\ref{table1} and Fig.~\ref{fig9}).These orientation dependent frequency peaks probably indicate a slight anisotropy of the corresponding electron pockets in BZ. The temperature dependence of the oscillating frequencies in Fig.~\ref{fig9} implies that the spin-orbit splitting probably increases with temperature.

\section{Acknowledgements}
The authors gratefully acknowledge the technical support of  V.A. Sidorov.

\end{document}